\documentclass[aps,prc,twocolumn,showpacs,superscriptaddress,nofootinbib]{revtex4-1}  
\usepackage{graphicx}  
\usepackage{dcolumn}   
\usepackage{bm}        
\usepackage{amssymb}   
\usepackage{amssymb,amsmath,amsfonts} 
\usepackage[breaklinks=true,debug=true]{hyperref}
\usepackage{braket}    
\usepackage{multirow}
\usepackage{adjustbox}
\usepackage{gnuplottex}
\usepackage[T1]{fontenc}
\usepackage[utf8]{inputenc}
\usepackage{color}
\usepackage{dsfont}
\usepackage{tikz}
\usepackage{filecontents}
\usepackage{pgfplots}

\usepgfplotslibrary{groupplots}
\pgfplotsset{compat=newest}
\usepgfplotslibrary{units}

\begin{document}


\title{Higher-order isospin-symmetry breaking corrections to nuclear matrix elements of superallowed $0^+\to 0^+$ Fermi $\beta$ decay of $T=1$ nuclei}

\author{L.~Xayavong} 
\affiliation{Physics Department, Faculty of Natural Sciences, National University of Laos, 7322 Dongdok, Vientiane Capital, Lao PDR}
\affiliation{CENBG (CNRS/IN2P3 -- Université de Bordeaux), 33175 Gradignan cedex, France}
\author{N.A.~Smirnova} 
\affiliation{CENBG (CNRS/IN2P3 -- Université de Bordeaux), 33175 Gradignan cedex, France}
\vskip 0.25cm  
\date{\today}

\begin{abstract} 

We study the shell-model formalism to include the isospin-symmetry-breaking correction ($\delta_{C}$) to nuclear matrix element of  superallowed $0^+\to 0^+$ Fermi $\beta$ decays of $T=1$ nuclei. Based on a perturbation expansion in a small quantity, such as the deviation of the overlap integral between proton and neutron radial wave functions from unity or of the transition density from its isospin-symmetry value, we derive that $\delta_C$ can be obtained as a sum of six terms, including two leading order (LO) terms, two next-to-leading order (NLO) terms, one next-to-next-to-leading order (NNLO) term and one next-to-next-to-next-to-leading order (NNNLO) term. The first two terms have been considered in a series of shell-model calculations of Towner and Hardy~\cite{ToHa2002, ToHa2008, HaTo2015x} as well as in the recent calculation of the present authors~\cite{XaNa2018,NaXa2018}, while the remaining four terms are usually neglected. A numerical calculation has been carried out for 13 transitions in the $p$, $sd$ and the lower part of $pf$ shells. 
Our result indicates that the magnitude of the sum of all higher order terms is of the order of $10^{-3}$~\%. 
This number is well below typical theoretical errors quantified within the shell model with 
Woods-Saxon radial wave functions~\cite{ToHa2008,XaNa2018}. 

\end{abstract}

\maketitle

\section{Introduction}

The Standard-Model description of the superallowed $0^+\to 0^+$ Fermi $\beta$ decay includes only the vector currents of the electroweak interaction. This important feature provides a very simple relationship between the vector-coupling constant, $G_V$ and the $ft$ value, 
with $\sim$1~\% theoretical corrections being applied to account for radiative effects and 
for isospin-symmetry breaking between a parent and a daughter state. 
It is customary to define a corrected $\mathcal{F}t$ value via 
\begin{equation}\label{Ft}
\displaystyle\mathcal{F}t = ft(1+\delta_R')(1-\delta_C+\delta_{NS}) = \frac{K}{|\mathcal{M}_F|^2G_V^2(1+\Delta_R^V)}, 
\end{equation}
where $ft$ is the product of the statistical rate function ($f$) and the partial half-life ($t$)~\cite{f}, 
$K$ is a combination of fundamental constants~\cite{OrBr1985}, 
$\delta_C$ is the correction for the breaking of the isospin symmetry which is the main interest of the present study~\cite{ToHa2008,XaNa2018}. The quantities $\Delta_R^V$, $\delta_R'$ and $\delta_{NS}$ are 
the nucleus-independent, the $(Z,Q_{EC})$-dependent, and the nuclear-structure-dependent radiative corrections, respectively~\cite{HaTo2020}. The Fermi matrix element in the isospin-symmetry limit is model independent and can be expressed as 
\begin{equation}
|\mathcal{M}_F|^2 = T(T+1)-T_{zi}T_{zf}, 
\end{equation}
where $T$ is the isospin quantum number of the multiplet, 
and $T_{zi}$ and $T_{zf}$ are the isospin projection quantum numbers of the initial and final nucleus, respectively. 
For an isospin triplet ($T=1$) we obtain $|\mathcal{M}_F|^2=2$ and hence $\mathcal{F}t$ should be a nucleus-independent quantity. 

As was discovered first by Cabibbo~\cite{Cabibbo1963}, and generalized further by Kobayashi and Maskawa~\cite{KobayashiMaskawa1973},
the vector-coupling constant, $G_V$, governing a semi-leptonic weak process 
is different from  the universal Fermi coupling constant, $G_\mu$, responsible for a purely leptonic weak decay~\cite{HaTo2020,Severijns2008}. 
The reason is that quarks participating in the weak interaction are superposition of the quark mass eigenstates.
This leads to appearence of the quark mixing matrix, or Cabbibo-Kobayashi-Maskawa (CKM) matrix, in the charge-changing
weak interaction Lagrangian. For a nucleon weak decay, 
\begin{equation}\label{ud}
G_V = G_\mu |V_{ud}|, 
\end{equation}
where $V_{ud}$ is the upper-left element of the CKM matrix. 

Therefore, precise determinations of $ft$ values together with theoretical corrections provide crucial information on the electroweak force and put constraints on physics beyond the Standard Model. 
For example, the constancy of $\mathcal{F}t$ values for all $J^\pi=0^+,T=1\to J^\pi=0^+,T=1$ decays would 
serve as a direct test of the Conserved Vector Current (CVC) hypothesis. 
The current average $\mathcal{F}t$ value for the 15 best-known superallowed transitions over a mass region of $10\le A\le 74$ is~\cite{HaTo2020}  
\begin{equation}
\overline{\mathcal{F}t} = 3072.24 \pm 1.85~\text{sec.},  
\end{equation}
with $\chi^2/\nu=0.47$. 

In addition, if CVC is confirmed, $|V_{ud}|$ can be extracted with a great precision from $\overline{\mathcal{F}t}$ via Eq.~\eqref{ud} and Eq.~\eqref{Ft}, which is important for the unitarity tests of the CKM matrix. 
Those tests would put stringint limits on possible physics beyond the Standard Model, such as the presence of scalar terms or right-handed currents. Further details and the current status of the domain can be found in Ref.~\cite{HaTo2020}. 

Although the $\delta_C$ correction is small, it is significant, and its associated theoretical errors, at present, dominate the uncertainty of $|V_{ud}|$ because of the very high precision reached on the experimental side and in the calculation of radiative corrections~\cite{HaTo2020}. Within the shell-model approach~\cite{ToHa2008,XaNa2018,OrBr1985}, $\delta_C$ is usually separated in two components, namely, 
\begin{equation}\label{C}
\delta_C\approx \delta_{C1} + \delta_{C2}, 
\end{equation}
where the first term on the right-hand side (r.h.s.) appears due to isospin-symmetry breaking effects 
in the configuration mixing induced by isospin-nonconserving forces in an effective shell-model Hamiltonian, 
whereas the second term accounts for a mismatch between neutron and proton single-particle radial wave functions.  

Calculations based on Eq.~\eqref{C} have provided the best set of $\delta_C$ values in eliminating the considerable scatters present within the uncorrected $ft$ values and, at the same time, excellently supported the top-row unitarity of the CKM matrix (see for example Ref.~\cite{ToHa2010,HaTo2020}). 
However, we recall that expression~\eqref{C} is only a lowest order approximation and its validity 
should be tested numerically for a large number of transitions. 
It is the purpose of the present study to derive a suitable formalism for missing higher order terms and 
to estimate their magnitude from a numerical calculation in the framework of the shell model with Woods-Saxon radial wave functions. 

The paper is structured as follows. In Section~\ref{sm} we present our theoretical formalism based on the shell model. 
Namely, starting from the basic definition of the Fermi $\beta$ decay matrix element within the closure approximation, 
we derive the isospin-symmetry-breaking corrections, 
including the LO and the higher order terms, and we discuss their properties. 
In the last part of this section, we generalize this idea and derive expressions for corrections within the parentage expansion formalism. In Section~\ref{com}, we present numerical calculations of the higher order terms and discuss their possible impact on the Fermi matrix element, 
as well as their relevance for the tests of the Standard Model. Conclusions and perspectives are given in Section~\ref{con}. 

\section{General shell-model formalism}\label{sm} 
\subsection{Closure approximation}

As a first step for deriving the exact shell-model expression of $\delta_C$, we write the nuclear matrix element for a Fermi transition from an initial state $\ket{i}$ to a final state $\ket{f}$ in the angular momentum coupled form, namely 
\begin{equation}\label{k}
M_F^\pm = \sum_{k_ak_b} \braket{k_a\tau_a||\tau_\pm||k_b\tau_b}\text{OBTD}(k_a\tau_ak_b\tau_bif\lambda), 
\end{equation} 
where $\ket{k_a\tau_a}\in\ket{f}$ and $\ket{k_b\tau_b}\in\ket{i}$ with $k_{a/b}$ standing for the set of spherical quantum number $(nlj)$ of state $a/b$ and $\tau_{a/b}$ for the isospin projection quantum number 
(we use the isospin convention of $\tau_p=-\frac{1}{2}$ for protons and $\tau_n=\frac{1}{2}$ for neutrons). 
The $\pm$ sign corresponds to the nuclear $\beta^\pm$ decay and $\tau_\pm$ is the isospin raising (upper sign)/lowering (lower sign) operator. 
The one-body transition density (OBTD) is defined as 
\begin{equation}\label{obtd}
\displaystyle\text{OBTD}(k_a\tau_ak_b\tau_bif\lambda) = \frac{ \braket{f|| [a_{k_b\tau_b}^\dagger \otimes \tilde{a}_{k_b\tau_b}]^\lambda ||i} }{ \sqrt{2\lambda+1} }, 
\end{equation}
where $\lambda=0(1)$ for Fermi (Gamow-Teller) $\beta$ decay. The double bars in the reduced matrix elements in Eq.~\eqref{k} and Eq.~\eqref{obtd} denote a reduction in angular momentum space. 

The reduced single-particle matrix element in Eq.~\eqref{k} can be written in a close form as follows 
\begin{equation}\label{kk}
\braket{k_a\tau_a||\tau_\pm||k_b\tau_b} = \theta_F(l_al_bj_aj_b) \Omega_{k_ak_b}^{\tau_a\tau_b} \xi_{\tau_a\tau_b},
\end{equation}
where the first factor, $\theta_F(l_al_bj_aj_b)$, depends on the orbital and total angular momenta of the single-particle states involved, and therefore specifies information on the transition's selection rule. 
For a Fermi operator, acting between the states of the same isospin multiplet ({\em isobaric analog states}), 
the function $\theta_F(l_al_bj_aj_b)$ has a very simple expression: 
\begin{equation}
\theta_{F}(l_al_bj_aj_b) = \sqrt{(2j_a+1)} \delta_{l_al_b}\delta_{j_aj_b}. 
\end{equation} 

The second factor on the r.h.s. of Eq.~\eqref{kk}, $\Omega_{k_ak_b}^{\tau_a\tau_b}$, is the overlap integral of 
single-particle radial wave functions: 
\begin{equation}\label{over}
\Omega_{k_ak_b}^{\tau_a\tau_b} = \int_0^\infty R_{k_a}^{\tau_a}(r)R_{k_b}^{\tau_a}(r) r^2 dr<1. 
\end{equation}

Note that if one uses harmonic oscillator functions which are isospin-invariant as employed in the conventional shell model, 
$\Omega_{k_ak_b}^{\tau_a\tau_b}$ reduces to the normalization integral. 
However, with realistic single-particle wave functions, such as the eigenfunctions of Woods-Saxon or Hartree-Fock potentials with Coulomb and nuclear isovector terms included, the integral $\Omega_{k_ak_b}^{\tau_a\tau_b}$ slightly deviates from unity. 
Furthermore, the inclusion of an isospin-nonconserving term in the mean-field potential leads to a nodal mixing in the eigenfunctions, 
and hence transitions between orbitals with different numbers of nodes are not strictly forbidden~\cite{MiSch2008, MiSch2009}. 
In general, this effect cannot be taken into account in a straightforward manner, because of the requirement of a huge model space. 

The isospin component, $\xi_{\tau_a\tau_b}$ of Eq.~\eqref{kk} is given by 
\begin{equation}
\xi_{\tau_a\tau_b} = \braket{\tau_a|\tau_\pm|\tau_b}  = \left\{
\begin{array}{ll}
1 & \text{for} \hspace{0.1in} \tau_b=\tau_a\mp 1, \\[0.18in]
0 & \text{otherwise.} 
\end{array}
\right.
\end{equation}

Within the framework of the shell model with realistic basis, isospin-symmetry breaking can impact the nuclear matrix element of the Fermi operator in two different ways: i) it creates differences in the structure of the initial and final states due to isospin mixing induced by isospin-nonconserving components of the effective shell-model Hamiltonian 
(this leads to the deviation of one-body transition densities from their isospin-symmetry-limit values), and
ii) it causes deviation of the overlap integrals~\eqref{over} from unity due to Coulomb and nuclear isovector terms present in the mean-field potential. Both effects lead to a reduction in absolute value of the Fermi matrix element~\cite{ToHa2008}. 

Therefore, it is natural to rearrange the $M_F^\pm$ expression as 
\begin{equation}\label{MF2}
\begin{array}{rl}
M_F^\pm = &\displaystyle  \sum_{k_ak_b} \theta_F(l_al_bj_aj_b)\xi_{\tau_a\tau_b} \text{OBTD}^T(k_a\tau_ak_b\tau_bif\lambda) \\[0.18in]
        - &\displaystyle  \sum_{k_ak_b} \theta_F(l_al_bj_aj_b) \Lambda_{k_ak_b}^{\tau_a\tau_b}\xi_{\tau_a\tau_b} \text{OBTD}^T(k_a\tau_ak_b\tau_bif\lambda) \\[0.18in]
        - &\displaystyle  \sum_{k_ak_b} \theta_F(l_al_bj_aj_b)\xi_{\tau_a\tau_b} D(k_a\tau_ak_b\tau_bif\lambda) \\[0.18in]
        + &\displaystyle  \sum_{k_ak_b} \theta_F(l_al_bj_aj_b) \Lambda_{k_ak_b}^{\tau_a\tau_b}\xi_{\tau_a\tau_b} D(k_a\tau_ak_b\tau_bif\lambda), 
\end{array}
\end{equation}
where $\text{OBTD}^T(k_a\tau_ak_b\tau_bif\lambda)$ stands for the isospin-symmetry limit of the one-body transition density and 
$D(k_a\tau_ak_b\tau_bif\lambda)$ for its deviation from the corresponding isospin-nonconserving value: 
\begin{equation}\label{D}
\begin{array}{ll}
D(k_a\tau_ak_b\tau_bif\lambda) &= \text{OBTD}^T(k_a\tau_ak_b\tau_bif\lambda) \\[0.18in] 
                               &-\text{OBTD}(k_a\tau_ak_b\tau_bif\lambda). 
\end{array}
\end{equation}

The quantity $\Lambda_{k_ak_b}^{\tau_a\tau_b}$ denotes the deviation from unity of the overlap integral, i.e.  
\begin{equation}\label{Lambda}
\Lambda_{k_ak_b}^{\tau_a\tau_b} = 1 - \Omega_{k_ak_b}^{\tau_a\tau_b}. 
\end{equation}

We remark that $\Lambda_{k_ak_b}^{\tau_a\tau_b}$ is always positive, 
while $D(k_ak_bif\lambda)$ can be either positive or negative.  In what follows we consider transitions between states for which isospin-symmetry is only weakly broken. In this case,  $\Lambda_{k_ak_b}^{\tau_a\tau_b}$ and $D(k_ak_bif\lambda)$ are sufficiently small quantities, so they can serve as perturbation parameters. 

The first term on the r.h.s. of Eq.~\eqref{MF2} corresponds to the Fermi matrix element in the isospin-symmetry limit, 
$\mathcal{M}_F^\pm $. We can therefore use it to factorize Eq.~\eqref{MF2} as  
\begin{equation}\label{MF3}
\begin{array}{rl}
M_F^\pm = &\displaystyle\mathcal{M}_F^\pm \Big[1 - \frac{1}{\mathcal{M}_F^\pm}\sum_{k_ak_b} \theta_F(l_al_bj_aj_b) \Lambda_{k_ak_b}^{\tau_a\tau_b} \xi_{\tau_a\tau_b} \\[0.18in] 
   \times &\displaystyle\text{OBTD}^T(k_a\tau_ak_b\tau_bif\lambda) \\[0.18in]
        - &\displaystyle\frac{1}{\mathcal{M}_F^\pm}\sum_{k_ak_b} \theta_F(l_al_bj_aj_b) \xi_{\tau_a\tau_b} D(k_a\tau_ak_b\tau_bif\lambda) \\[0.18in]
        + &\displaystyle\frac{1}{\mathcal{M}_F^\pm}\sum_{k_ak_b} \theta_F(l_al_bj_aj_b) \Lambda_{k_ak_b}^{\tau_a\tau_b} \xi_{\tau_a\tau_b} D(k_a\tau_ak_b\tau_bif\lambda) \Big].  
\end{array}
\end{equation}

The last three terms on the r.h.s. of Eq.~\eqref{MF3} appear due to isospin-nonconservation. 
If isospin symmetry is preserved, those terms vanish and, hence, $M_F^\pm = \mathcal{M}_F^\pm$.

At the next step, we square both sides of Eq.~\eqref{MF3} and rearrange the result in the following form
\begin{equation}\label{MF4}
|M_F^\pm|^2 = |\mathcal{M}_F^\pm|^2(1-\delta_C), 
\end{equation}
where the total isospin-symmetry-breaking correction, $\delta_C$, represents a sum of six terms: 
\begin{equation}\label{cc}
\delta_C = \overline{\delta}_{C1} + \overline{\delta}_{C2} + \overline{\delta}_{C3} + \overline{\delta}_{C4} + \overline{\delta}_{C5} + \overline{\delta}_{C6}, 
\end{equation}
here the bar indicates that the correction terms are evaluated within the closure approximation. A more extended treatment is present in subsection~\ref{parent}. 

The first and second terms on the r.h.s. of Eq.~\eqref{cc} are the two usual LO terms. It is interesting to note that, at this lowest order approximation, the isospin mixing and the radial mismatch effects can be accounted for as two separate correction terms. 
The isospin-mixing correction corresponds to $\overline{\delta}_{C1}$. 
This correction term is calculated using an isospin-nonconserving effective shell-model Hamiltonian and the harmonic oscillator basis, such as
\begin{equation} \label{LO}
\begin{array}{ll}
\overline{\delta}_{C1} &= \displaystyle\frac{2}{\mathcal{M}_F^\pm}\sum_{k_ak_b} \theta_F(l_al_bj_aj_b)\xi_{\tau_a\tau_b} D(k_a\tau_ak_b\tau_bif\lambda), \\[0.18in]
                 &= \displaystyle 2-\frac{2}{\mathcal{M}_F^\pm}\sum_{k_ak_b} \theta_F(l_al_bj_aj_b)\xi_{\tau_a\tau_b} \\[0.18in]
                 & \times \text{OBTD}(k_a\tau_ak_b\tau_bif\lambda). 
\end{array}
\end{equation}

The radial mismatch correction corresponds to $\overline{\delta}_{C2}$. This correction term is calculated using an isoscalar effective shell-model Hamiltonian and realistic radial wave functions, namely
\begin{equation}
\begin{array}{ll}
\overline{\delta}_{C2} & = \displaystyle\frac{2}{\mathcal{M}_F^\pm}\sum_{k_ak_b} \theta_F(l_al_bj_aj_b) \Lambda_{k_ak_b}^{\tau_a\tau_b} \xi_{\tau_a\tau_b} \\[0.18in] 
  & \times \text{OBTD}^T(k_a\tau_ak_b\tau_bif\lambda), \\[0.18in]
  & = \displaystyle 2 -\frac{2}{\mathcal{M}_F^\pm}\sum_{k_ak_b} \theta_F(l_al_bj_aj_b) \Omega_{k_ak_b}^{\tau_a\tau_b} \xi_{\tau_a\tau_b} \\[0.18in] 
  & \times \text{OBTD}^T(k_a\tau_ak_b\tau_bif\lambda). 
\end{array}
\end{equation}

The third and the fourth terms on the r.h.s. of Eq.~\eqref{cc} are the NLO terms. The former depends on both $\Lambda_{k_ak_b}^{\tau_a\tau_b}$ and $D(k_a\tau_ak_b\tau_bif\lambda)$, therefore it must be evaluated using both an isospin non-conserving Hamiltonian and realistic radial wave functions. It can be expressed as
\begin{equation} \label{NLO}
\begin{array}{ll}
\overline{\delta}_{C3}&=\displaystyle-\frac{2}{\mathcal{M}_F^\pm}\sum_{k_ak_b} \theta_F(l_al_bj_aj_b) \Lambda_{k_ak_b}^{\tau_a\tau_b} \xi_{\tau_a\tau_b} D(k_a\tau_ak_b\tau_bif\lambda), \\[0.18in]
&= \displaystyle-\overline{\delta}_{C2}+\frac{2}{\mathcal{M}_F^\pm}\sum_{k_ak_b} \theta_F(l_al_bj_aj_b) \Lambda_{k_ak_b}^{\tau_a\tau_b} \xi_{\tau_a\tau_b} \\[0.18in]
& \times \text{OBTD}(k_a\tau_ak_b\tau_bif\lambda).
\end{array}
\end{equation}

In contrast, the latter is simply a function of the two LO terms, which can be written as 
\begin{equation} 
\overline{\delta}_{C4}=\displaystyle -\frac{ \left(\overline{\delta}_{C1} + \overline{\delta}_{C2}\right)^2  }{4}. 
\end{equation}

The fifth and the sixth terms on the r.h.s. of Eq.~\eqref{cc} are the NNLO and NNNLO terms, respectively. 
$\overline{\delta}_{C5}$ is determined by the two NLO terms, 
\begin{equation} \label{NNLO}
\overline{\delta}_{C5} = \displaystyle - \frac{ ( \overline{\delta}_{C1} + \overline{\delta}_{C2} )\overline{\delta}_{C3} }{2} = -\overline{\delta}_{C3}\sqrt{|\overline{\delta}_{C4}|}, 
\end{equation}
while $\overline{\delta}_{C6}$ is determined only by $\overline{\delta}_{C3}$ as
\begin{equation}
\overline{\delta}_{C6} = \displaystyle -\frac{(\overline{\delta}_{C3})^2}{4}. 
\end{equation}

Apparently, one only needs to perform shell-model calculations for the first three terms of Eq.~\eqref{cc} because the other three terms are just combinations of them. It can be also noticed that the LO terms are generally positive as can be seen from the previous calculations~\cite{ToHa2008,XaNa2018}, $\overline{\delta}_{C4}$ and $\overline{\delta}_{C6}$ are obviously negative, $\overline{\delta}_{C3}$ can be negative or positive, while the sign of $\overline{\delta}_{C5}$ is opposite to the sign of $\overline{\delta}_{C3}$.  

\subsection{Parentage expansion formalism}\label{parent}

The proton and neutron single-particle wave functions used for the evaluation of the overlap integrals depend on the type and parameterization of the realistic single-particle potential. 
For example, Towner and Hardy~\cite{ToHa2008} worked mainly with a phenomenological Woods-Saxon potential, 
whereas Ormand and Brown~\cite{OrBr1985} employed a local equivalent potential based on a self-consistent Hartree-Fock calculation 
with an effective zero-range Skyrme interaction. In both cases, the chosen potential was thoroughly re-adjusted 
so that the energy eigenvalues would match the experimental separation energies. 
This procedure ensures the robustness of radial wave functions in the asymptotic region, as is clear from the following equation
$$
\displaystyle R(r) \to \exp{\left( -\frac{\sqrt{2m |\epsilon |}r}{\hbar}\right)},
$$
with $\epsilon$ and $m$ denoting the single-particle energy and the nucleon mass, respectively.  

In order to specify separation energies needed to constraint the potential depth we insert a complete set of states $\ket{\pi}$ of the $(A-1)$-nucleon system into the one-body transition densities in Eq.~\eqref{obtd} between the creation and annihilation operators. As a result, $\delta_{C2}$ takes the form    
\begin{equation} \label{delta}
\begin{array}{ll}
{\delta}_{C2} = & \displaystyle\frac{2}{\mathcal{M}_F^\pm}\sum_{k_ak_b\pi} \theta_F(l_al_bj_aj_b) \Lambda_{k_ak_b}^{\tau_a\tau_b\pi} \xi_{\tau_a\tau_b} \\[0.18in]
& \displaystyle \times \Theta(j_aj_bJ_iJ_fJ_\pi\lambda) A^T(f;\pi k_a\tau_a)A^T(i;\pi k_b\tau_b),  
\end{array}
\end{equation}
where $A^T(f;\pi k_a\tau_a)$ and $A^T(i;\pi k_b\tau_b)$ stand for the spectroscopic amplitudes obtained from an  
isoscalar effective shell-model Hamiltonian. They are defined as 
\begin{equation}\label{af}
A^T(f;\pi k_a\tau_a) =\frac{ (f|| a_{k_a\tau_a}^\dagger ||\pi) }{ \sqrt{2J_f+1} }, 
\end{equation}
and 
\begin{equation}\label{ai}
A^T(i;\pi k_b\tau_b) =\frac{ (i|| a_{k_b\tau_b}^\dagger ||\pi) }{ \sqrt{2J_i+1} }, 
\end{equation}
where $J_i$ and $J_f$ are angular momenta of the initial and final states, respectively. 

Again, double bars in Eq.~\eqref{af} and Eq.~\eqref{ai} denote reduction in angular momentum space. 
It should be also noted that we use round brackets for an isospin-invariant many-particle state. 
$\Lambda_{k_ak_b}^{\tau_a\tau_b}$ in Eq.~\eqref{delta} contains an additional label $\pi$, 
indicating that it is evaluated with radial wave functions
whose asymptotic form matches separation energies with respect to excited states of the $(A-1)$-nucleon system. 
More details can be found in Ref.~\cite{ToHa2008}.  
 
The function $\Theta(j_aj_bJ_iJ_fJ_\pi\lambda)$ appearing in Eq.~\eqref{delta} is given by
\begin{equation}
\begin{array}{ll}
\Theta(j_aj_bJ_iJ_fJ_\pi\lambda) &= \sqrt{(2J_i+1)(2J_f+1)} \\[0.18in]
 & \times (-1)^{J_f+J_\pi+j_a+\lambda} 
\left\{
\begin{array}{lll}
J_i & J_f & \lambda \\
j_b & j_a & J_\pi
\end{array}
\right\}, 
\end{array}
\end{equation}
where $J_\pi$ is the angular momentum of the intermediate state $\ket{\pi}$.  

In the same way, the expression of $\delta_{C3}$ is evaluated as  
\begin{equation} \label{deltaC3}
\begin{array}{ll}
\delta_{C3} = & \displaystyle-\delta_{C2}+\frac{2}{\mathcal{M}_F^\pm}\sum_{k_ak_b\pi} \theta_F(l_al_bj_aj_b) \Lambda_{k_ak_b}^{\tau_a\tau_b\pi} \xi_{\tau_a\tau_b} \\[0.18in] 
& \displaystyle \times \Theta(j_aj_bJ_iJ_fJ_\pi\lambda) A(f;\pi k_a\tau_a)A(i;\pi k_b\tau_b).
\end{array}
\end{equation}

It can be remarked here that the structure of the second term on the r.h.s. of Eq.~\eqref{deltaC3} looks very similar to that of $\delta_{C2}$, except that $A^T(f;\pi k_a\tau_a)$ and $A^T(i;\pi k_b\tau_b)$ are replaced with the spectroscopic amplitudes calculated using an isospin non-conserving shell-model Hamiltonian (without superscript $T$). 
Furthermore, $\delta_{C3}$ will be negative if this term is smaller than $\delta_{C2}$ and positive in the opposite case.  

In contrast, the isospin-mixing correction (the first term on the r.h.s. of Eq.~\eqref{cc}) is not affected by this expansion because it does not depend on radial wave functions. Therefore, we can write 
\begin{equation} 
\begin{array}{ll}
{\delta}_{C1} = & \displaystyle \overline{\delta}_{C1}. 
\end{array}
\end{equation}

All the other correction terms must be re-evaluated, taking into account the parentage expansion. 
In partiular, $\delta_{C4}$ becomes 
\begin{equation} 
{\delta}_{C4} =  \displaystyle -\frac{ \left( {\delta}_{C1} + {\delta}_{C2} \right)^2 }{4}, 
\end{equation}
the new expression for $\delta_{C5}$ reads
\begin{equation}
{\delta}_{C5} =  \displaystyle -\frac{ ({\delta}_{C1} + {\delta}_{C2}){\delta}_{C3} }{2}, 
\end{equation}
and similarly for $\delta_{C6}$ we have 
\begin{equation}
{\delta}_{C6} =  \displaystyle -\frac{ ({\delta}_{C3})^2}{4}. 
\end{equation}

We notice that calculations in the full parentage expansion formalism consume much more computational resources 
than calculations in the closure approximation. 
In general, one hundred of intermediate states of each spin and parity must be included, otherwise the corrections would not converge. 
The numerical aspects of the calculation of $\delta_{C2}$ for the $sd$-shell emitters have been discussed in Ref.~\cite{XaNa2018}.

\begin{table*}[ht!]
\caption{\label{tab1} Calculated values of various terms of the isospin-symmetry-breaking correction in percent unit. Here LO, NLO, NNLO and NNNLO are the abbreviations for leading order, next-to-leading order, next-to-next-leading order and next-to-next-to-next-leading order, respectively.}
\begin{ruledtabular}
\begin{tabular}{c|c|c|c|c|c|c|c}
\multirow{2}{*}{Emitter} &     \multicolumn{2}{c|}{LO}                 &   \multicolumn{2}{c|}{NLO}                    &   NNLO & NNNLO & NLO+NNLO+NNNLO \\
\cline{2-8}
&$\delta_{C1}$	&$\delta_{C2}$	&$\delta_{C3}$ &$\delta_{C4}$	&$\delta_{C5}$	&$\delta_{C6}$	& $\delta_{C3}+\delta_{C4}+\delta_{C5}+\delta_{C6}$	\\ 
\hline
 $^{10}$C &  0.03421939 &  0.18931000 &  0.00049200 & -0.00012491 & -0.00000055 & -0.00000000 &  0.00036654 \\
 $^{14}$O &  0.01016209 &  0.28316400 & -0.00234000 & -0.00021510 &  0.00000343 & -0.00000001 & -0.00255168 \\
$^{18}$Ne &  0.00796504 &  0.20549300 &  0.00377500 & -0.00011391 & -0.00000403 & -0.00000004 &  0.00365702 \\
$^{22}$Mg &  0.01987732 &  0.26357900 &  0.00073900 & -0.00020087 & -0.00000105 & -0.00000000 &  0.00053708 \\
$^{26}$Al &  0.00793319 &  0.26326700 &  0.00021200 & -0.00018387 & -0.00000029 & -0.00000000 &  0.00002784 \\
$^{26}$Si &  0.03037112 &  0.36937000 &  0.00058500 & -0.00039948 & -0.00000117 & -0.00000000 &  0.00018435 \\
 $^{30}$S &  0.05890865 &  0.68247200 &  0.00514100 & -0.00137411 & -0.00001906 & -0.00000007 &  0.00374776 \\
$^{34}$Cl &  0.04336312 &  0.61015300 & -0.00112700 & -0.00106771 &  0.00000368 & -0.00000000 & -0.00219103 \\
$^{34}$Ar &  0.00913932 &  0.70812600 & -0.00055800 & -0.00128617 &  0.00000200 & -0.00000000 & -0.00184217 \\
 $^{42}$Ti &  0.00577545 &  0.37658000 &  0.00068900 & -0.00036549 & -0.00000132 & -0.00000000 &  0.00032219 \\
  $^{46}$V &  0.03266025 &  0.34878700 &  0.00029300 & -0.00036376 & -0.00000056 & -0.00000000 & -0.00007131 \\
 $^{46}$Cr &  0.02236702 &  0.44804100 &  0.00068200 & -0.00055321 & -0.00000160 & -0.00000000 &  0.00012719 \\
$^{50}$Mn &  0.04100000 &  0.46533600 & -0.00076300 & -0.00064094 &  0.00000193 & -0.00000000 & -0.00140201 \\
\end{tabular}
\end{ruledtabular}
\end{table*}

\section{Numerical calculation of the higher order terms}\label{com}  

Within the shell-model formalism discussed in the previous section, we have carried out a numerical calculation of the higher order terms for 13 superallowed $0^+\to 0^+$ nuclear $\beta$ decays, including $^{10}$C, $^{14}$O, $^{18}$Ne, $^{22}$Mg, $^{26m}$Al, $^{26}$Si, $^{30}$S, $^{34}$Cl, $^{34}$Ar, $^{42}$Ti, $^{46}$V, $^{46}$Cr and $^{50}$Mn. We selected the Cohen-Kurath interaction~\cite{Cohen1967} for nuclei with mass between $10$ and $14$, the well-known universal $sd$-shell interaction of Wildenthal~\cite{USD} for nuclei in the range of $18\le A\le 34$, and the so-called GXPF1A interaction of Honma and collaborators~\cite{gx1a} for those with $42\le A\le 50$. The respective configuration spaces are the full $p$, $sd$ and $pf$ shells. 
The isospin non-conserving counterpart of the above cited shell-model effective Hamiltonians is comprised of isovector single-particle energies, the two-body Coulomb force between protons and phenomenological charge-dependent nucleon-nucleon potentials of nuclear origin. 
Details of the fitting procedure are described in Ref.~\cite{OrBr1989x}. Our large-scale diagonalizations have been performed using the NuShellX@MSU~\cite{NuShellX} shell-model code. 

It can be noticed that in most cases, our chosen model spaces are smaller than those used in the calculations by Towner and Hardy~\cite{ToHa2008}. 
We are aware that these reduced model spaces might not produce all necessary configurations for the initial and final states of the decays under consideration. Nevertheless, they should be sufficient for our present study which aims at exploring a relative magnitude of various subleading terms of the isospin-symmetry-breaking correction. 

The overlap integrals were evaluated with eigenfunctions of a phenomenological Woods-Saxon potential 
with the parametrization of Bohr and Mottelson~\cite{BohrMott}, supplemented by modifications as described in Ref.~\cite{XaNa2018}. 
In particular, the potential depth has been re-adjusted case-by-case in order to reproduce experimental separation energies, 
while accounting for excitations of the intermediate $(A-1)$-nucleon system. 
In addition, the Woods-Saxon length parameter has been simultaneously re-adjusted to reproduce the measured value of the charge radius of the emitter. Note that, for a given transition, we have kept the length parameter the same for the initial and final nuclei. More details on the parameter adjustment, including our formalism for the charge radius calculation can be found in Ref.~\cite{XaNa2018}. 

For the reason of consistency, we did not use the existing values of $\delta_{C1}$ and $\delta_{C2}$, but we have recalculated them on equal footing with $\delta_{C3}$. For the same reason, we did not scale $\delta_{C1}$ with the energy separation between the analogue and the nearest anti-analogue states in daughter nuclei as suggested by Towner and Hardy~\cite{ToHa2008}. Moreover, since we are interested only in the relative magnitude between various correction terms, it is not necessary to consider uncertainties from the use
of several effective shell-model Hamiltonians and from the charge radius data, which can be quantified using the method described in Refs.~\cite{Xthesis,ToHa2002,ToHa2008}.  

Our results for all correction terms are listed in Table~\ref{tab1}. 
Although the theoretical analysis in the previous section supposes that $\delta_{C1}$ and $\delta_{C2}$ are of the same order of magnitude, the calculated $\delta_{C1}$ values are generally considerably smaller than $\delta_{C2}$. 
It is clearly seen that, for most cases, our calculated values for the two LO terms differ significantly from those of Towner and Hardy~\cite{ToHa2008}, the reason is related to the difference in configuration spaces, effective Hamiltonians, the Woods-Saxon parametrization and the potential-adjustment procedure. 

It is interesting to remark that the sign of $\delta_{C3}$ varies from transition to transition as expected from the theoretical inspection in the previous section. We obtained a negative $\delta_{C3}$ value for $^{14}$O, $^{34}$Cl, $^{34}$Ar and $^{50}$Mn, while a positive value for the others. The principal reason is that the isospin-mixing effect on the transition densities is non-monotonic -- it can be larger or smaller, depending on both the nucleus and the orbitals involved. The sign of the other correction terms can be obtained from the following consideration. 
According to our numerical results, the absolute value of $\delta_{C3}$ is, on average, one order of magnitude smaller than $\delta_{C1}$ and two orders of magnitude smaller than $\delta_{C2}$. We also see that $\delta_{C4}$ is, on average, of the same order of magnitude as $\delta_{C3}$, but its sign is always negative, therefore a cancellation between these NLO terms can happen in some cases. At the same time, $\delta_{C5}$ and $\delta_{C6}$ are completely negligible. 

Finally, we obtain that the sum of all higher order terms for a given transition (see the last column of Table~\ref{tab1}), 
is generally smaller or, at most, comparable to the uncertainties on the LO terms published in Ref.~\cite{ToHa2008}. 
Therefore, their presence would not produce any significant impact on the existing values of $\delta_C$, and can be neglected for the tests of the Standard Model. It does not seem that the higher order terms would increase dramatically with mass number. Nevertheless, they are determined by the magnitude of the LO terms, especially by $\delta_{C2}$, which is very sensitive to the weakly-bound effect~\cite{ToHa2008,OrBr1985,XaNa2018}. 
Therefore, the higher order terms may be significant only if the LO terms are found to be abnormally large. 

\section{Conclusion and perspective}\label{con} 
We have developed a shell-model formalism for exact calculation of isospin-symmetry-breaking correction to superallowed $0^+\to 0^+$ nuclear $\beta$ decay of $T=1$ nuclei. Our special attention has been focused on the higher order terms of this correction 
which were not considered by any of the previous shell-model calculations. 
Our numerical results indicate that contributions of the higher order terms are smaller or, at most, comparable with typical uncertainties on the lowest-order correction terms. 
However, the magnitude of higher-order terms would increase with increasing magnitude of the perturbation parameters
(i.e. deviation of the overlap integrals from unity or the deviation of the one-body transition densities from their isospin-symmetry-limit values). Thus, higher order terms, especially NLO terms, may be significant in the cases when the LO terms are extremely large, i.e. the superallowed $0^+\to 0^+$ nuclear $\beta$ decays with $T=2$~\cite{Bhattacharya} and some other cases as discussed in Ref.~\cite{Kaneko}. The theoretical formalism derived in this article can be easily generalized for other nuclear weak processes, such as Fermi $\beta$ decay of $J\ne 0,T\ne 1$ multiplets and Gamow-Teller $\beta$ decay. 

\begin{acknowledgments} 
We are grateful to B.~Blank for a careful reading of the manuscript. 
The work is supported by IN2P3/CNRS, France, in the framework of the ``Isospin-symmetry breaking'' and 
``Exotic nuclei, fundamental interactions and astrophysics'' Master projects. 
\end{acknowledgments}

\bibliography{higher_order_isospin_corrections}

\end{document}